\documentclass[prb,twocolumn,superscriptaddress,showpacs,%
preprintnumbers,amsmath,amssymb]{revtex4}
\usepackage{amsmath}
\usepackage{graphicx}
\usepackage{dcolumn}

\begin{document}

\title{Wave packet dynamics and valley filter in strained graphene}

\author{Andrey Chaves}\email{andrey@fisica.ufc.br}
\affiliation{Department of Physics, University of Antwerp,
Groenenborgerlaan 171, B-2020 Antwerp,
Belgium}\affiliation{Departamento de F\'isica, Universidade
Federal do Cear\'a, Caixa Postal 6030, Campus do Pici, 60455-900
Fortaleza, Cear\'a, Brazil}
\author{L. Covaci}\affiliation{Department of Physics, University of
Antwerp, Groenenborgerlaan 171, B-2020 Antwerp, Belgium}
\author{Kh. Yu. Rakhimov}\affiliation{Department of Physics, University of
Antwerp, Groenenborgerlaan 171, B-2020 Antwerp,
Belgium}\affiliation{Heat Physics Department of the Uzbek Academy
of Sciences, 100135 Tashkent, Uzbekistan}
\author{G. A. Farias}\email{gil@fisica.ufc.br}
\affiliation{Departamento de F\'isica, Universidade Federal do
Cear\'a, Caixa Postal 6030, Campus do Pici, 60455-900 Fortaleza,
Cear\'a, Brazil}
\author{F. M. Peeters}\email{francois.peeters@ua.ac.be}\affiliation{Department of Physics, University of
Antwerp, Groenenborgerlaan 171, B-2020 Antwerp,
Belgium}\affiliation{Departamento de F\'isica, Universidade
Federal do Cear\'a, Caixa Postal 6030, Campus do Pici, 60455-900
Fortaleza, Cear\'a, Brazil}

\date{ \today }

\begin{abstract}
The time evolution of a wavepacket in strained graphene is studied
within the tight-binding model and continuum model. The effect of
an external magnetic field, as well as a strain-induced
pseudo-magnetic field, on the wave packet trajectories and
zitterbewegung are analyzed. Combining the effects of strain with
those of an external magnetic field produces an effective magnetic
field which is large in one of the Dirac cones, but can be
practically zero in the other. We construct an efficient valley
filter, where for a propagating incoming wave packet consisting of
momenta around the $K$ and $K'$ Dirac points, the outgoing wave
packet exhibits momenta in only one of these Dirac points, while
the components of the packet that belong to the other Dirac point
are reflected due to the Lorentz force. We also found that the
zitterbewegung is permanent in time in the presence of either
external or strain-induced magnetic fields, but when both the
external and strain-induced magnetic fields are present, the
zitterbewegung is transient in one of the Dirac cones, whereas in
the other cone the wave packet exhibits permanent spatial
oscillations.
\end{abstract}

\pacs{72.80.Vp,73.23.-b, 85.75.Hh}

\maketitle

\section{Introduction}

Since its first synthesis in 2004, \cite{Novoselov} graphene has
been attracting much interest due to its unique electronic
properties arising from its singular energy spectrum, where in the
vicinity of the points labelled as $K$ and $K'$ in reciprocal
space, the charge carriers behave as massless quasi-particles and
exhibit an almost linear dispersion. \cite{CastroNeto} These
quasi-particles obey the Dirac-Weyl equations and therefore should
be subject to quasi-relativistic effects, such as zitterbewegung,
i.e., a trembling motion caused by interference between positive
and negative energy states. \cite{Thaller, Zawadzki1, David} The
phenomenon of zitterbewegung was predicted in 1930 by
Schr\"odinger \cite{Schrodinger} and has been subject of renewed
interest over the past years. Previous theoretical works have
suggested few ways of experimentally observing zitterbewegung, e.
g. in narrow-gap semiconductors \cite{Zawadzki}, in III-V
zinc-blende semiconductor quantum wells with spin-orbit coupling
\cite{Schliemann} and, more recently, in monolayer \cite{Rusin1}
and bilayer \cite{Wang} graphene. An experimental simulation of
the zitterbewegung of free relativistic electrons in vacuum was
performed by Gerritsma et al. \cite{Gerritsma} by using trapped
ions.

Strain engineering in graphene has recently become a widely
studied topic. \cite{PereiraECastroNeto1, Choi, Pellegrino,
Treidel, Fujita, Chang} The elastic properties of graphene
nanoribbons were theoretically investigated by Cadelano \emph{et
al.}\cite{Cadelano}, which studied the in-plane stretching and out
of plane bending deformations by combining continuum elasticity
theory and tight-binding atomistic simulations. Later, Cocco
\emph{et al.}\cite{Cocco} and Lu and Guo \cite{Lu} showed that a
combination of shear and uniaxial strain at moderate absolute
deformations can be used to open a gap in the graphene energy
spectrum. It has been shown recently that specific forms of strain
produce a pseudo-magnetic field in graphene, which does not break
the time reversal symmetry and which points in opposite directions
for electrons moving around the $K$ and $K'$ points.
\cite{Vozmediano} The strain-induced magnetic field is expected to
produce Landau levels and, consequently, the quantum Hall effect,
even in the absence of an external magnetic field. \cite{Guinea,
Low} Guinea \emph{et al.}, \cite{Guinea2} showed theoretically
that an in-plane bending of the graphene sheet leads to an almost
uniform field. Landau levels as a consequence of strain-induced
pseudo-magnetic fields greater than 300 Tesla were recently
observed with scanning tunnelling microscopy (STM) in nanometer
size nanobubbles. \cite{Levy}

Although previous works have studied wave packet propagation for
the Dirac-Weyl Hamiltonian of graphene in the absence of external
fields and potentials, \cite{Maksimova} or in the presence of
uniformly applied external magnetic fields \cite{Rusin1, Rusin2, Romera,
Krueckl}, there is still a lack of theoretical works on the wave
packet propagation through potential and (pseudo) magnetic field
step barriers. Moreover, the time evolution of a wave packet in
graphene within the tight-binding (TB) model, where the
intra-valley scattering to higher energy states and inter-valley
scattering due to defects appear naturally, is still hardly found
in the literature. It is also interesting to see whether the
results from Dirac and TB approaches for graphene differ or are
similar.

In this paper, we investigate the time-evolution
of wavepackets in graphene within the TB model, based on the
split-operator technique for the expansion of the time-evolution
operator. We trace a parallel between the results from the TB
model and those obtained from the Dirac approximation for particles
with momentum close to one of the Dirac cones of the Brillouin
zone of graphene. The proposed method is then applied to the study
of the dynamics of Gaussian wave packets in graphene under
external magnetic fields. The effects of the pseudo-magnetic field
induced by bending the graphene sheet into an arc of circle are
analyzed as well. Our results show that for an appropriate choice
of strain and external magnetic field strength, the system
exhibits a strong effective magnetic field for particles in one of
the Dirac cones, whereas in the other cone the external and
pseudo-magnetic fields cancel each other and the effective magnetic
field is practically zero. We show that this effect is manifested
as a transient (permanent) zitterbewegung for electrons in the
cone where the effective magnetic field is zero (non-zero), which
can be verified experimentally by detecting the electric field
radiation emitted by the trembling wave packet. \cite{Rusin2}
Moreover, our results show that with such a choice of external and
strain-induced magnetic fields, one can construct an efficient
valley-filter, which can be useful for future valley-tronic
devices. \cite{Rycerz}

\section{Time evolution operator}

By solving the time-dependent Schr\"odinger equation, one obtains
that the propagated wavefunction after a time step $\Delta t$ can
be calculated by applying the time-evolution operator on the wave
packet at any instant $t$
\begin{equation}\label{eqTimeEvolutionOp}
\Psi (\vec{r},t + \Delta
t) = e^{-\frac{i}{\hbar}H \Delta t}\Psi(\vec{r},t),
\end{equation}
where we assumed that the Hamiltonian $H$ is time-independent.
Different techniques have been suggested for the expansion of the
exponential in Eq. (\ref{eqTimeEvolutionOp}), e.g. the Chebyschev
polynomials method \cite{Fehske} and the second order differencing
scheme \cite{Mahapatra, Kosloff}. The numerical method that we use
for the application of the time evolution operator in this work,
namely, the split-operator method, \cite{Chaves} is the subject of
this section, where we will show how this technique can be adapted
for the study of the wave packet dynamics in graphene within the
tight-binding and Dirac approximations.

\subsection{Tight-Binding model}

Graphene consists of a layer of carbon atoms forming a honeycomb
lattice, which can be described by the Hamiltonian
\begin{equation} \label{TBHamiltonian}
H_{TB} =
\sum_{i}\epsilon_ic^{\dagger}_ic_i+\sum_{<i,j>}\left[\tau_{ij}c^{\dagger}_ic_j+\tau^*_{ij}c_ic_j^{\dagger}\right],
\end{equation}
where $c_i$($c_i^{\dagger}$) annihilates (creates) an electron at
the site $i$, with on-site energy $\epsilon_i$, and the sum is
taken only between nearest neighbor sites $i$ and $j$, with
hopping energy $\tau_{ij}$. The effect of an external magnetic
field can be calculated by including a phase in the hopping
parameters according to the Peierls substitution $\tau_{ij}
\rightarrow \tau_{ij}\exp\left[i\frac{e}{\hbar}\int_j^i\vec{A}
\cdot d\vec{l}\right]$, where $\vec{A}$ is the vector potential
describing the magnetic field. \cite{Bahamon, Governale} In a
strained graphene sheet, the distance between two adjacent sites
$i$ and $j$ is changed by $\Delta a_{ij} = a_{ij} - a_0$, where
$a_0$ is the lattice parameter of unstrained graphene and $a_{ij}$
is the distance between the sites after the stress. The change in
the inter-sites distance affects the hopping energy between the
sites, which becomes \cite{Vozmediano} $\tau_{ij} \rightarrow
\tau_{ij}\left(1 + 2\Delta a_{ij}/a_0\right)$. A similar
expression can be obtained by expanding Eq. (13) of Ref.
\onlinecite{PereiraECastroNeto} in Taylor series and neglecting
higher order terms in $\Delta a_{ij}$, i.e., considering small
lattice deformations. The strain-induced change in the hopping
energies leads to an effective pseudo-magnetic field, which points
to opposite directions in the valley $K$ and $K'$. \cite{Guinea}
Notice that the pseudo-magnetic field in our model is not
introduced artificially by considering an additional vector
potential in the Peierls phase, but it rather appears naturally
after the changes in the inter-site distances due to the strain.

\begin{figure}[!bpht]
\centerline{\includegraphics[width=\linewidth]{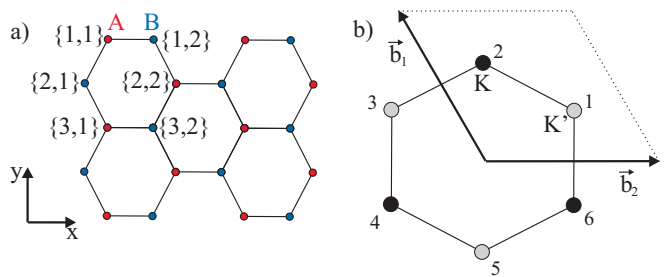}}
\caption{(Color online) (a) Sketch of the honeycomb lattice of
graphene, made out of two superimposed triangular lattices $A$ and
$B$. The atoms are labelled as $\{n,m\}$ according to their line
and column numbers $n$ and $m$, respectively. (b) Reciprocal
lattice of graphene, with $K$ (black) and $K'$ (gray) Dirac
points, where the area defined by the reciprocal vectors
$\vec{b}_1 = (-2\pi/3a_0,\sqrt{3})$ and $\vec{b}_2 =
(4\pi/3a_0,0)$ represents the first Brillouin zone. The numbers
close to each Dirac point are explained in the
text.}\label{fig:Honeycomb}
\end{figure}

Let us label the sites $i$ of the graphene lattice according to
their line and column numbers $n$ and $m$, respectively, as shown
in Fig. \ref{fig:Honeycomb}(a). The basis vector state
$|n,m\rangle$ represents an electron confined on the site of line
$n$ and column $m$. In a lattice consisting only of
non-interacting sites, each $|n,m\rangle$ is an eigenstate of
$H_{TB}$ with energy $\epsilon_{n,m}$, i.e. $H_{TB}|n,m\rangle =
\epsilon_{n,m}|n,m\rangle$. Limiting ourselves to nearest neighbors
interactions, we find
\begin{eqnarray} \label{HTBfull}
H_{TB}|n,m\rangle = \epsilon_{nm}|n,m\rangle +
T_{m+1}|n,m+1\rangle \quad\quad\quad \nonumber\\
+ T_{m-1}|n,m-1\rangle +
T_{n+1}|n+1,m\rangle + T_{n-1}|n-1,m\rangle,
\end{eqnarray}
where $T_{n\pm1}$ and $T_{m\pm1}$ are equivalent to the hopping energies $\tau_{ij}$
between the site $i = \{n,m\}$ and the adjacent sites $j = \{n\pm1,m\}$ and
$j = \{n,m\pm1\}$, respectively.
Eq. (\ref{HTBfull}) can be rewritten as
\begin{equation}\label{HTBbest}
H_{TB}|n,m\rangle = H_n|n,m\rangle + H_m|n,m\rangle,
\end{equation}
where the operators $H_n$ and $H_m$ are defined as
\begin{subequations}\label{HTBbroken}
\begin{equation}
H_{n}|n,m\rangle = T_{m+1}|n,m+1\rangle + T_{m-1}|n,m-1\rangle +
\frac{\epsilon_{nm}}{2}
\end{equation}
and
\begin{equation}
H_{m}|n,m\rangle = T_{n+1}|n+1,m\rangle + T_{n-1}|n-1,m\rangle +
\frac{\epsilon_{nm}}{2}.
\end{equation}
\end{subequations}
The wavefunction at any instant $t$ is
then written as a linear combination of the basis vector states
$\Psi^{t}_{n,m} = \sum_{n,m} b^{t}_{n,m}|n,m\rangle$. The
advantage of following the procedure described by Eqs.
(\ref{HTBfull}-\ref{HTBbroken}) lies in the fact that the
operators $H_n$ and $H_m$ in Eq. (\ref{HTBbroken}) can be
represented by tri-diagonal matrices, which are easier to handle
than the matrix representing the full Hamiltonian, Eq.
(\ref{HTBfull}).

The split-operator technique can now be applied to the Hamiltonian
Eq. (\ref{HTBbest}), so that the time evolution operator is
approximated by
\begin{equation}
e^{-\frac{i}{\hbar}H_{TB}\Delta t}=
e^{-\frac{i}{2\hbar}H_{m}\Delta t}e^{-\frac{i}{\hbar}H_{n}\Delta
t}e^{-\frac{i}{2\hbar}H_{m}\Delta t} + O(\Delta t^3),
\end{equation}
where the error comes from the non-commutativity between the
operators $H_n$ and $H_m$. We drop the $O(\Delta t^3)$ terms by
considering a small time step $\Delta t = 0.1$ fs. The propagated
wavefunction is then obtained from Eq. (\ref{eqTimeEvolutionOp}),
which in this case reads
\begin{equation}\label{EqTBsplitted}
\Psi^{t+\Delta t}_{n,m} = e^{-\frac{i}{2\hbar}H_{m}\Delta
t}e^{-\frac{i}{\hbar}H_{n}\Delta t}e^{-\frac{i}{2\hbar}H_{m}\Delta
t}\Psi^{t}_{n,m}.
\end{equation}
This equation is solved in three steps:
\begin{subequations}\label{splitoperatorHTB}
\begin{equation}
\eta_{n,m} = e^{-\frac{i}{2\hbar}H_{m}\Delta t}\Psi^t_{n,m},
\end{equation}
\begin{equation}
\xi_{n,m} = e^{-\frac{i}{\hbar}H_{n}\Delta t}\eta_{n,m},
\end{equation}
\begin{equation}
\Psi^{t+\Delta t}_{n,m} = e^{-\frac{i}{2\hbar}H_{m}\Delta
t}\xi_{n,m}.
\end{equation}
\end{subequations}
Using the Cayley form for the exponentials, \cite{Watanabe} we can
rewrite Eq. (\ref{splitoperatorHTB}a) as
\begin{eqnarray}
\eta_{n,m} = e^{-\frac{i}{2\hbar}H_{m}\Delta t}\Psi^{t}_{n,m} =
\frac{1-\frac{i\Delta t}{4\hbar}H_{m}}{1+\frac{i\Delta
t}{4\hbar}H_{m}}\Psi^{t}_{n,m} + O (\Delta t^2),
\end{eqnarray}
which leads to
\begin{eqnarray}\label{cayley}
\left(1+\frac{i\Delta t}{4\hbar}H_{m}\right)\eta_{n,m} \approx
\left(1-\frac{i\Delta t}{4\hbar}H_{m}\right)\Psi^{t}_{n,m}.
\end{eqnarray}
As the wavefunction $\Psi^{t}_{n,m}$ is already known, the matrix
equation in Eq. (\ref{cayley}) can be straightforwardly solved to
obtain $\eta_{n,m}$. We repeat this procedure for the other two
exponentials in Eqs. (\ref{splitoperatorHTB}b) and
(\ref{splitoperatorHTB}c), and eventually obtain $\Psi^{t+\Delta
t}_{n,m}$.

In fact, the form in Eq. (\ref{cayley}) can also be applied to the
full Hamiltonian $H_{TB}$, i.e. without splitting the $H_m$ and
$H_n$ terms. However, this would lead to matrix equations
involving penta-diagonal matrices, which are harder to handle than
the tri-diagonal matrices in Eq. (\ref{cayley}). As the error
produced by the splitting in Eq. (\ref{EqTBsplitted}) is smaller
than the error produced by the (necessary) expansion of the
exponential given by Eq. (\ref{cayley}), it is worthy to split
these terms in order to simplify the numerical calculations.

\subsection{Dirac-Weyl equation}

From the TB Hamiltonian Eq. (\ref{TBHamiltonian}), considering an
infinite graphene sheet in the absence of external potential and
magnetic fields, one obtains the energy bandstructure of graphene
as
\begin{equation}\label{bandstructure}
E\left(\vec{k}\right) = \pm \tau\sqrt{3+f\left(\vec{k}\right)}
\nonumber
\end{equation}
\begin{equation}
f\left(\vec{k}\right) = 2\cos\left(\sqrt{3}k_ya_0\right) +
4\cos\left(\frac{\sqrt{3}}{2}k_ya_0\right)\cos\left(\frac{3}{2}k_xa_0\right),
\end{equation}
which is gapless in six points of the reciprocal space where $E =
0$, from which only two are inequivalent, labelled as $K$ and
$K'$, as shown in Fig. \ref{fig:Honeycomb}(b).
\cite{CastroNeto, PereiraReview} In the vicinity of each of these points, the
dependence of the energy spectrum on the wave vector $\vec{k}$ is
almost linear and the electron can be described as a massless
fermion by the Dirac Hamiltonian
\begin{equation} \label{DiracHamiltonian}
H_D = \left[v_F\vec{\sigma}\cdot(\vec{\textsf{p}}+e\vec{A}) +
V(\textsf{x},\textsf{y})\textbf{I}\right]e^{-i\phi} ,
\end{equation}
where $v_F = 3\tau a_0/2\hbar$ is the Fermi velocity, $\vec{A}$ is
the electromagnetic vector potential, $V(x,y)$ is an external
potential, $\textbf{I}$ is the identity matrix, $\vec{\sigma}$ is the
Pauli vector and the wavefunctions are pseudo-spinors $\Psi =
[\Psi_A, \Psi_B]^T$, with $\Psi_{A(B)}$ as the probability of
finding the electron in the sub-lattice $A (B)$ that composes the
honeycomb lattice of graphene. \cite{CastroNeto} The angle $\phi$
is different for electrons around the $K$ and $K'$ Dirac cones. In the vicinity
of the $k$-th Dirac point (see labels for each Dirac point in Fig.
\ref{fig:Honeycomb}(b)), one obtains $\phi = -\pi/6 + k\pi/3$,
with $k = 1-6$. From here onwards, we will refer to the
coordinates in the Dirac (TB) model as $\textsf{x}$ ($x$) and
$\textsf{y}$ ($y$).

The exponential term in Eq. (\ref{DiracHamiltonian}) is usually
dropped, because it can be considered as a phase
in the state vectors in the Dirac model. However, this term has an
important meaning when comparing with the TB model: this
exponential can be identified as a rotation operator with
angle $\phi$. Notice that an infinite graphene hexagonal lattice
has C$_{6v}$ symmetry, i.e. it is symmetric only for rotation
angles $k\pi/3$ ($k$ - integer). As a consequence, the Dirac
Hamiltonian Eq. (\ref{DiracHamiltonian}) without the exponential
term is not symmetric in the momenta $\hat{p}_x$ and
$\hat{p}_y$, as already pointed out previously. \cite{Rusin1} So,
what would be the meaning of the direction-dependent observables
in the Dirac description of graphene, when they are not symmetric
under rotation, exhibiting a privileged direction? Defining $y$ ($x$)
as the zigzag (armchair) direction, as in Fig.
\ref{fig:Honeycomb}(a), the results obtained by the Dirac
approximation for the $\textsf{x}$ and $\textsf{y}$ components of
any observable are compared to the armchair and zigzag directions,
respectively, after performing a rotation $\phi$ in the
coordinates of the Dirac model. From the possible values of
$\phi$, one deduces straightforwardly that at any Dirac cone, the
coordinate $\textsf{x}$ ($\textsf{y}$) of the Dirac model is
always related to one of the zigzag (armchair) directions of the
real graphene lattice. On the other hand, for finite rectangular
samples the different angles $\phi$ represent two distinguishable
situations, since the rectangle does not share the C$_{6v}$
symmetry of the infinite graphene lattice: the $\textsf{x}$
direction in the Dirac model for the odd (even) cones in Fig.
\ref{fig:Honeycomb}(b) represents a diagonal (vertical) direction
in the rectangle. The comparison between the TB model for a
rectangular graphene flake and the Dirac approximation will be
discussed in details further, in the following section.

In a recent work, Maksimova \textit{et al.} \cite{Maksimova}
presented an analytical study of the time evolution of a Gaussian
wave packet in graphene in the \textit{absence} of external potentials
and/or magnetic fields within the continuum model, i. e. using the
Dirac-Weyl Hamiltonian for electrons in the vicinity of the Dirac point
$K$. In this paper, we will use an alternative and more general
way of calculating the dynamics of a wavepacket in such a system,
\cite{PereiraReview} based on the split-operator technique, which
can be applied for systems under arbitrary external potentials and
magnetic fields.

The Dirac-Weyl Hamiltonian $H_D$ in Eq. (\ref{DiracHamiltonian})
can be separated as $H_D = H_k + H_r$, where $H_k = \hbar
v_F\vec{\sigma}\cdot\vec{\textsf{k}}$ keeps only the terms which
depend on the wave vector $\vec{\textsf{k}}$, whereas $H_r =
v_Fe\vec{\sigma}\cdot\vec{A} + V\textbf{I}$ depends only on the
real space coordinates $\textsf{x}$ and $\textsf{y}$. Following
the split-operator method, the time evolution operator for the
Hamiltonian $H_D$ can be approximated as
\begin{eqnarray}
\exp\left[-\frac{i\Delta t}{\hbar}\left(H_k + H_r\right)\right]
\approx \nonumber \quad\quad\quad\quad
\\
\exp\left[-\frac{i \Delta t}{2\hbar}H_r\right]\exp\left[-\frac{i
\Delta t}{\hbar}H_k\right]\exp\left[-\frac{i \Delta
t}{2\hbar}H_r\right]
\end{eqnarray}
with an error of the order of $O(\Delta t^3)$, due to the
non-commutativity of the operators involved. We invoke a
well-known property of the Pauli vectors
\begin{equation}
\exp\left(-i\vec{S}\cdot\vec{\sigma}\right) =
\cos(S)\textbf{I}-i\frac{\sin(S)}{S}\left(\vec{S}\cdot\vec{\sigma}\right)
\end{equation}
for any vector $\vec{S}$, where $S = |\vec{S}|$, and rewrite the
exponentials in real and reciprocal space, respectively, in
matrix form
\begin{subequations}
\begin{equation}
\textbf{M}_r = \left[\cos\left(\textsf{A} \right)\textbf{I} -i \frac{\sin\left(\textsf{A} \right)}{\textsf{A}}\left(%
\begin{array}{cc}
   0 & \textsf{A}_\textsf{x}-i\textsf{A}_\textsf{y} \\
 \textsf{A}_\textsf{x}+i\textsf{A}_\textsf{y} & 0 \\
\end{array}%
\right)\right]e^{-\frac{i\Delta t}{2\hbar}V}
\end{equation}
\begin{equation}
\textbf{M}_k = \cos(\kappa)\textbf{I} -i\frac{\sin(\kappa)}{\kappa}\left(%
\begin{array}{cc}
  0 & \kappa_x-i\kappa_y \\
  \kappa_x+i\kappa_y & 0 \\
\end{array}%
\right),
\end{equation}
\end{subequations}
where $\textsf{A} = |\vec{\textsf{A}}| = \Delta t v_Fe|\vec{A}|\big/2\hbar$, $\vec{\kappa} = \Delta t v_F\vec{\textsf{k}}$ and $\kappa =
|\vec{\kappa}|=\Delta t
v_F\sqrt{\textsf{k}_\textsf{x}^2+\textsf{k}_\textsf{y}^2}$, so
that the time evolution of a wave packet $\Psi_D(x,y) = [\phi_A,
\phi_B]^T\Psi(x,y)$ can be calculated as a series of matrix
multiplications
\begin{equation}
\Psi(\vec{\textsf{r}},t+\Delta t) =
\textbf{M}_r\cdot\textbf{M}_k\cdot\textbf{M}_r\Psi(\vec{\textsf{r}},t)
+ O(\Delta t^3).
\end{equation}
The matrix multiplication by $\textbf{M}_k$ is made in reciprocal
space by taking the Fourier transform of the functions. In the
absence of magnetic fields and external potentials, one has
\textbf{M}$_r$ = \textbf{I} and
\begin{equation}
\Psi(\vec{\textsf{r}},t+\Delta t)
=\textbf{M}_k\Psi(\vec{\textsf{r}},t),
\end{equation}
where the matrix multiplication in reciprocal space gives the
exact result for the time evolution of the wave packet, since
there is no error induced by non-commutativity of operators or
matrices in this case. This shows that the split-operator method
provides a way to study the dynamics of wavepackets in graphene
within the continuum model where, in the presence of magnetic
fields and/or external potentials, one can control the accuracy of
the results by making $\Delta t$ smaller, while in their absence,
the problem is solved exactly by a simple matrix multiplication
for any value of $\Delta t$.

\section{Results and discussion}

We shall now discuss the results obtained for a graphene lattice
with 2000$\times$3601 atoms, with armchair (zigzag) edges in the
$x$($y$)-direction. The nearest neighbors hopping parameter and
the lattice constant of graphene are taken as $\tau = -2.7$ eV and
$a_0 = 1.42$ \AA\,, respectively.

As initial wave packet, we consider a Gaussian centered at
$\vec{r_0} = (x_0, y_0)$ in real space and $\vec{q} =
(q_x^0,q_y^0)$ in reciprocal space:
\begin{equation}\label{initial}
\Psi_q(\vec{r}) = N\left(%
\begin{array}{c}
  c_1 \\
  c_2 \\
\end{array}
\right)\exp\left[-\frac{(x-x_0)^2+(y-y_0)^2}{2d ^2}
+ i \vec{q}\cdot\vec{r}\right],
\end{equation}
where $N$ is the normalization factor. Notice that we have
included a pseudo-spinor $[c_1, c_2]^T$ in the initial wave
packet, where $c_{1(2)}$ is the probability of finding the
electron in the triangular sub-lattice $A (B)$ that composes the
graphene hexagonal lattice. One can also rewrite the pseudo-spinor
as $[1, e^{i\theta}]^T$, where the pseudo-spin polarization angle
$\theta$ is shown explicitly. The pseudo-spin is a concept
normally attributed to the Dirac description of graphene. Indeed,
the pseudo-spin of a wavefunction in the Dirac model is related to
the expectation values of the Pauli matrices
$\langle\sigma_i\rangle$, which can involve integrals of the
product between wavefunctions for sub-lattices $A$ and $B$. Such a
definition fails for the TB wavefunctions, since in this case they
are defined in different points of the lattice, so that any
integral that mixes functions of both sub-lattices gives zero. Even so, the
study of the pseudo-spin related to the initial discrete wave
packet helps to understand the wave packet dynamics obtained from
the TB model, as we will see further in this section.

\subsection{Initial pseudo-spin polarization and zitterbewegung revisited}

In this subsection, we will use the TB model to review some of the
main properties of the wave packet dynamics in graphene. Within
the TB model, we consider the initial wave packet as a discrete
form of the Gaussian distribution in Eq. (\ref{initial}) for the
graphene hexagonal lattice, where we multiply the Gaussian
function by $c_{1(2)}$ in the sites belonging to the triangular
sub-lattice $A (B)$. From Eq. (\ref{bandstructure}), it is clear
that in momentum space, the region of interest is the vicinity
of the Dirac points $K$ and $K'$, since the energy corresponding
to wave vectors out of this region is very high.
\begin{figure}[!bpht]
\centerline{\includegraphics[width=\linewidth]{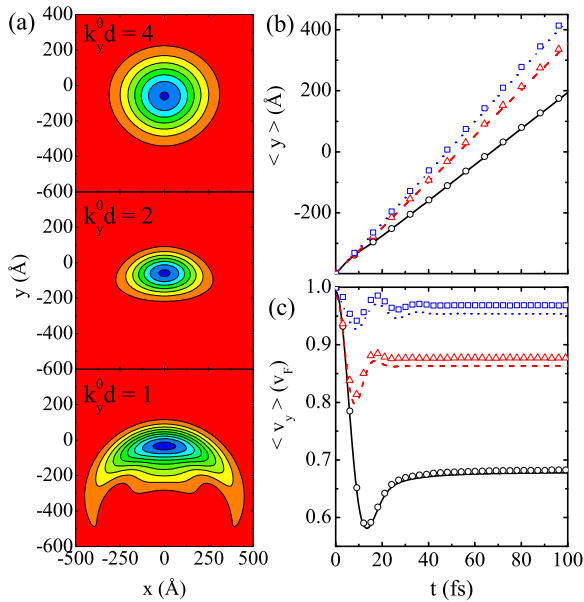}}
\caption{(Color online)(a) Contour plots of the squared modulus of
the wavefunction after a time evolution of $t = 40$ fs, for three
different values of the dimensionless parameter $k_y^0d$. (b)
Expectation value of the position and (c) velocity in
$y$-direction as a function of time. The results obtained from the
TB (Dirac) model are presented as curves (symbols), for $k_y^0d =
1$ (solid, circles), 2 (dashed, triangles) and 4 (dotted, squares)
}\label{fig:timeposition}
\end{figure}

In the TB model for two-dimensional crystals, one usually
considers the same Gaussian distribution for all the sites of the
lattice. \cite{Nazareno} This is equivalent as choosing $c_1 = c_2
= 1$ in Eq. (\ref{initial}). Figure \ref{fig:timeposition}(a)
shows the contour plots of the squared modulus of the propagated
wavefunctions at $t = 40$ fs for these values of $c_i$,
considering an initial wave vector $\vec{q} = (0,k_y^0) + K$, i.e.
in the vicinity of the $K$ point labelled as 2 in Fig.
\ref{fig:Honeycomb}(b). As shown in Ref. \onlinecite{Maksimova},
the wave packet dynamics near the Dirac cones in graphene does not
depend separately on the momentum $k_y^0$ or on the width $d$, but
rather on the dimensionless quantity $k_y^0d$. This result was
obtained from the Dirac model for graphene, i.e. considering that
even high energies states exhibit linear dispersion. Within the TB
model we expect that wave packets with the same $k_y^0d$ behave
alike only if $k_y^0$ is not too far from the Dirac cone and if
$d$ is not too small, so that the packet is well localized in
energy space. Within these conditions, Fig. \ref{fig:timeposition}
shows the time evolution for different values of this
dimensionless quantity: $k_y^0d = 1$ and 2, with $d = 100$ \AA\,,
and $k_y^0d = 4$, with $d = 200$ \AA\,. We observe that the
dispersion of the wave packet is stronger for smaller values of
$k_y^0d$, where it becomes distorted into an arc-like shape. For
larger $k_y^0d$, on the other hand, the wave packet keeps its
circularly symmetric shape for longer times.
\begin{figure}[!b]
\centerline{\includegraphics[width=0.8\linewidth]{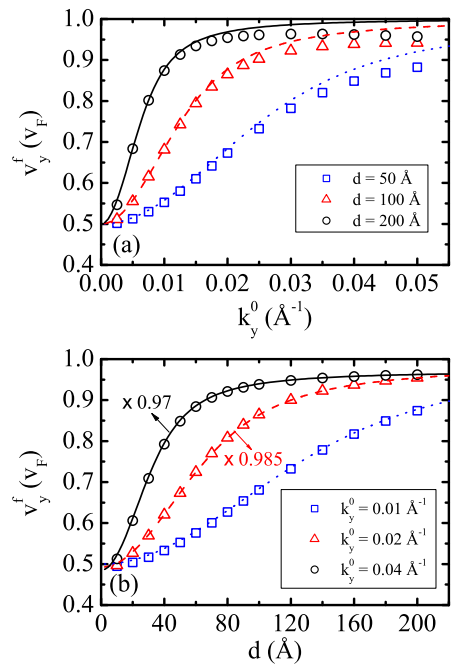}}
\caption{(Color online) Final velocities for the Gaussian wave
packet in Eq. (\ref{initial}), with pseudo-spin $c_1 = c_2 = 1$
and momentum $\vec{q} = (0,k_y^0) + K$ as a function of (a) the
momentum $k_y^0$, for widths $d = 50$ \AA\,, 100 \AA\, and 200
\AA\,, and (b) the width $d$, for momenta $k_y^0 = 0.01$
\AA\,$^{-1}$, 0.02 \AA\,$^{-1}$ and 0.04 \AA\,$^{-1}$. The symbols
(curves) are obtained from the TB (Dirac) model. In (b), the
results from the Dirac model for $k_y^0 =$ 0.02 \AA\,$^{-1}$
(dashed) and 0.04 \AA\,$^{-1}$ (solid) are multiplied by 0.985 and
0.97, respectively.}\label{fig:velocityVSdANDk}
\end{figure}

As explained in the previous subsection, in order to obtain the
Dirac Hamiltonian Eq. (\ref{DiracHamiltonian}), one must shift the
origin of the wave vectors $\vec{k}$ to one of the six Dirac
points shown in Fig. \ref{fig:Honeycomb}(b). Besides, one must
also rotate the $x$ and $y$ axis by an angle $\phi$ which depends
on the $K$ or $K'$ point that is considered as the origin in
momentum space. For the $K = (0,4\pi\big/3\sqrt{3}a_0)$ point,
labelled as 2 in Fig. \ref{fig:Honeycomb}(b), the Dirac
Hamiltonian Eq. (\ref{DiracHamiltonian}) is obtained by rotating
the axis by 90$^{\circ}$, with other words, by a transformation of
the coordinates as $x \rightarrow -\textsf{y}$ and $y \rightarrow
\textsf{x}$. The pseudo-spinor $c_1 = c_2 = 1$ represents a wave
packet polarized in $\textsf{x}$-direction, i.e. $\langle
\sigma_x \rangle > 0$ and $\langle \sigma_z \rangle = \langle
\sigma_y \rangle = 0$. From the Heisenberg picture, we obtain the
velocity in the $\textsf{x}$-direction for the proposed wave packet as
\begin{equation}
\frac{d \textsf{x}}{dt} =
\frac{1}{i\hbar}\left[\textsf{x},H_D\right] = v_F \sigma_x.
\end{equation}
Performing the appropriate coordinate transformations, the
velocity obtained from the TB model for the $y$-direction must be
consistent with the prediction from the Dirac approximation,
namely, $v_y = d\textsf{x}/dt = v_F\sigma_x$. This suggests that
such a wave packet propagates towards the positive $y$-direction,
but with non-constant velocity, since $\sigma_x$ does not commute
with $H_D$. The expectation value of the $y$ position of the
packet $\langle y \rangle$ is shown as a function of time by the
curves in Fig. \ref{fig:timeposition}(b), for $k_y^0d = 1$
(solid), 2 (dashed) and 4 (dotted), where the results obtained by
the Dirac equation are shown as symbols for comparison. A
different linear behavior is already observed for each wave packet
at large time, implying that they have different velocities, which
is kind of counter-intuitive, since low-energy electrons in
graphene are expected to propagate always with the same Fermi
velocity $v_F$. Figure \ref{fig:timeposition}(c) shows the
velocity $v_y$, calculated by taking the derivative of the TB
results for $\langle y \rangle$ with respect to time, which
exhibits clear oscillations that are damped as time evolves,
converging to a final value $v^f_y < v_F$ that depends on the
initial wave packet's width $d$ and wave vector $k_y^0$. The
velocities obtained by the Dirac model are shown by symbols, where
the same qualitative behavior is observed as obtained from the TB
model, though for higher wave packet momentum and width, a small
quantitative difference is observed, which is a consequence of the
different energy-momentum dispersion.
\begin{figure}[!b]
\centerline{\includegraphics[width=\linewidth]{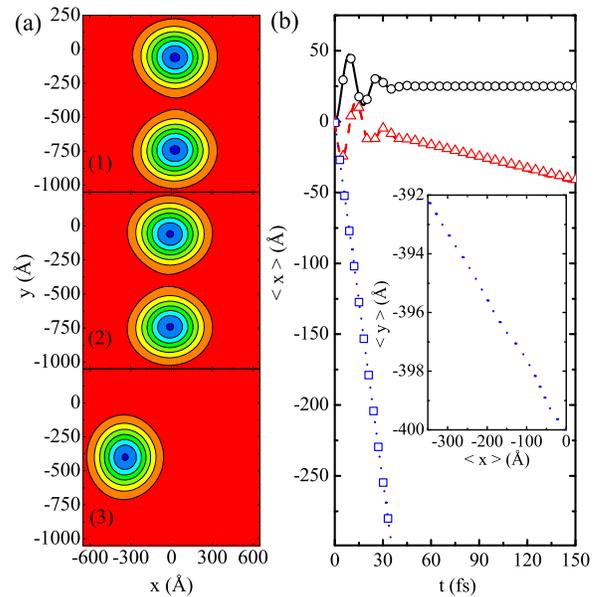}}
\caption{(Color online)(a) Contour plots of the squared modulus of
the wavefunction after a time evolution of $t = 40$ fs, for three
initial configurations of pseudo-spin $[c_1,c_2]^T$ and momentum
$\vec{q}_0$: 1) $[1,0]^T$, $k_x^0 = 0$ and $k_y^0d = 4$; 2)
$[1,i]^T$, $k_x^0 = 0$ and $k_y^0d = 4$; and 3) $[1,i]^T$, $k_x^0d
= 4$, $k_y^0 = 0$. (b) Expectation value of $x$ obtained by the TB
model for the initial wavepackets 1 (solid), 2 (dashed) and 3
(dotted) as a function of time. The results obtained by the Dirac
model, after the appropriate coordinates rotation (see text), are
shown as circles, triangles and squares, respectively. The inset
shows the trajectory of the wavepacket obtained from the TB model
for the initial wavepacket 3. }\label{fig:timeposition10}
\end{figure}

The oscillatory behavior of the velocity is a
manifestation of the zitterbewegung, i.e. a trembling motion of
the wave packet due to the interference between positive and
negative energy states that makes up the initial Gaussian wave
packet. \cite{Rusin1, Thaller} This effect is well-known for
relativistic particles, which are described by the Dirac equation,
and is also observed for electrons in graphene in the vicinity of
the $K$ and $K'$ points, since they can be described as massless
quasi-particles by the Dirac equation as well. The velocity
wiggles with shorter period and smaller amplitudes for larger
values of $k_y^0d$. The convergence of the velocities demonstrates
that the zitterbewegung is not a permanent, but a transient
effect. \cite{Rusin1}

Figure \ref{fig:velocityVSdANDk} shows the converged velocity
$v^f_y$ as a function of (a) the momentum $k^0_y$ and (b) the
width $d$ of the Gaussian wave packet. The TB results (symbols)
are compared to those calculated from Eq. (31) in
Ref. \onlinecite{Maksimova} (curves), which was obtained
analytically from the Dirac approximation in the $t \rightarrow
\infty$ limit and is repeated here just for completeness:
\begin{equation}\label{eqCompleteness}
\frac{v^f_y}{v_F} = 1-\frac{1-e^{-(k^0_yd)^2}}{2(k^0_yd)^2}.
\end{equation}
Within the Dirac model, one can observe that increasing $d$ or
$k_y^0$ in Eq. (\ref{eqCompleteness}), the final velocity
increases monotonically and approaches $v_F$, which is reasonable,
since a wider packet in real space leads to a narrower
distribution in $k$-space, whereas a higher value of the wave
vector makes the center of the packet lay far from $E = 0$. In
both cases, the interference with negative energy states is
reduced and, consequently, zitterbewegung effects are less
significant. However, since the analytical formula Eq.
(\ref{eqCompleteness}) does not take into account any effect such
as the curvature of the energy bands for higher energy states or
trigonal warping effects, this formula is not expected to give
accurate results for larger $k^0_y$. Indeed, Fig.
\ref{fig:velocityVSdANDk}(a) shows that a very good agreement
between TB and Dirac results can be observed only for small values
of $k^0_y$, whereas for large $k^0_y$, the final velocities
obtained from the TB model are lower than those obtained from the
Dirac model and do not increase monotonically, but decreases
slowly for very large $k^0_y$, as a consequence of the curvature
of the energy bands. On the other hand, in Fig.
\ref{fig:velocityVSdANDk}(b) we observe that varying the wave
packet width for a fixed momentum, good \emph{qualitative}
agreement with the Dirac model is obtained for almost any value of
$d$. The curves for larger values of $k^0_y$ (solid and dashed)
are just \emph{quantitatively} different from those obtained by
the TB model, and they are comparable to the TB results after
multiplication by a factor 0.985 (0.970) for $k^0_y = 0.02$
\AA\,$^{-1}$ (0.04 \AA\,$^{-1}$). Worse qualitative agreement
between TB and Dirac results in this case is observed only for
very small $d$, where the Gaussian width in energy space, given by
$\Delta E = v_F\hbar d^{-1}$, incorporates higher energy values,
leading to deviations in $v^f_y$ obtained from the TB model as
compared to those from the Dirac model.
\begin{figure}[!t]
\centerline{\includegraphics[width=\linewidth]{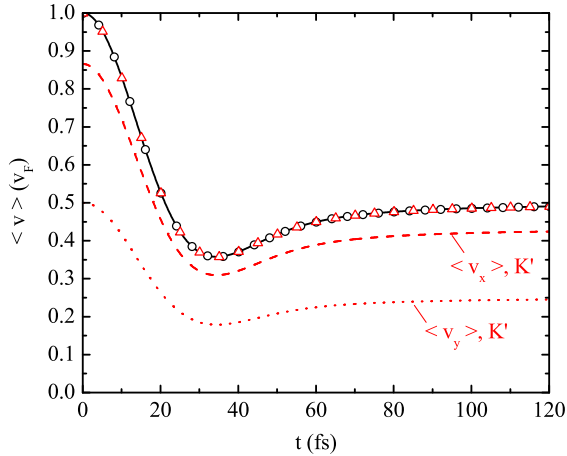}}
\caption{(Color online)(a) Expectation value of the velocity as a
function of time, for wavepackets with $k_y^0 = k_x^0 = 0$ and
pseudo-spinor $[1,1]^T$ (solid) and $[1,i]^T$ (circles) at the
Dirac point $K = (0,4\pi\big/3\sqrt{3}a_0)$ (point 2 in Fig.
\ref{fig:Honeycomb}(b)), and $[1,1]^T$ (triangles) at
$K'=(2\pi\big/3a_0,2\pi\big/3\sqrt{3}a_0)$ (point 1 in Fig.
\ref{fig:Honeycomb}(b)). The $x$ and $y$ components of the
velocity in the latter case are shown as dashed and dotted curves,
respectively.}\label{fig:velKKprime}
\end{figure}

In Fig. \ref{fig:timeposition10}(a) we show contour plots of the
squared modulus of the wave function at $t = 40$ fs for three
different choices of wave vectors $\vec{q} = (k_x^0,k_y^0) + K$
and initial pseudo-spinors: 1) $[1,0]^T$, with $k_x^0 = 0$ and
$k_y^0d = 4$, 2) $[1,i]^T$, with $k_x^0 = 0$ and $k_y^0d = 4$, and
3) $[1,i]^T$, with $k_x^0d = 4$ and $k_y^0 = 0$. The curves
(symbols) in Fig. \ref{fig:timeposition10}(b) show the expectation
value $\langle x \rangle$ for each case obtained by the TB (Dirac)
model. In case 1 (solid, circles), the pseudo-spinor points in the
$z$-direction, so that $\langle \sigma_x \rangle = \langle
\sigma_y \rangle$ and, consequently, the velocity for both
in-plane directions must be zero. Indeed, the wave packet splits
into two equal parts that propagate in opposite $y$ directions,
leading to $v_y = 0$. In the $x$-direction, although there is
still a small zitterbewegung, $\langle x \rangle$ rapidly
converges to a constant, leading to $v_x = 0$. In case 2 (dashed,
triangles), the pseudo-spinor points in the
$\textsf{y}$-direction, but the momentum of the wave packet in
this direction is zero, so that the packet splits in the
$y$-direction, since $v_y = v_F\sigma_x = 0$, but drifts slowly in
the $-x$ direction (or, equivalently, $\textsf{y}$). In case 3
(dotted, squares), both the pseudo-spin and the momentum are in
the $\textsf{y}$-direction, so that the wave packet propagates in
this direction without any splitting. This situation is comparable
to the one in Fig. \ref{fig:timeposition}(a), since in both cases
the pseudo-spin and momentum are in the same direction and, as a
consequence, the wave packet propagates in this direction
practically preserving its circular symmetry. However, in the case
3, the packet still presents a very weak oscillation in the
$y$-direction, and also drifts very slowly in this direction, as
one can see from the trajectory of the packet in the $x-y$ plane
for this case, shown in the inset of Fig.
\ref{fig:timeposition10}(b). This oscillation and drift are
related to the contributions of higher energy states in the wave
packet: a wave packet centered around $k_x^{(0)} = 0$ and
$k_y^{(0)} \neq 0$, as in Fig. \ref{fig:timeposition}(a), has a
symmetric distribution of momenta in $x$-direction even for higher
energies and, consequently, there is no additional oscillation in
this direction. On the other hand, a packet centered around
$k_x^{(0)} \neq 0$ and $k_y^{(0)} = 0$, as in Fig.
\ref{fig:timeposition10}(b), does not have a symmetric
distribution of momenta in the $y$-direction due to the trigonal
warping for higher energies and, consequently, some oscillations
are observed in this direction. As the standard Dirac Hamiltonian
$H_D$ for graphene does not take trigonal warping into account,
this effect is not observed in the Dirac model.

In the numerical work of Thaller, \cite{Thaller} as well as in the
analytical work of Maksimova, \cite{Maksimova} it is demonstrated
within the Dirac model that even when $k_y^0 = k_x^0 = 0$, wave packet
motion is still observed due to zitterbewegung effects. The
velocities of the wave packet obtained from our TB model of
graphene for wave packet momenta exactly at the $K'$ and $K$, i.e.
points 1 and 2 in Fig. \ref{fig:Honeycomb}(a), respectively, are shown in Fig.
\ref{fig:velKKprime}. The velocities exhibit a damped oscillation with
the same time-dependent modulus for any pseudo-spin and Dirac
point, though they point in different directions: for $[1,1]^T$
(solid) and $[1,i]^T$ (circles) in $K$, the velocity is in the $y$
and $-x$ directions of the lattice, respectively, which are
exactly the directions of polarization of these pseudo-spinors
after the $\phi = \pi/2$ rotation required by the $K$ cone 2. In
the $K'$ cone 1, the rotation angle is $\phi = \pi/6$ and,
accordingly, the velocity points in this direction, as one can see
by the decomposition of the velocity in the components $\langle
v_x \rangle$ (dashed) and $\langle v_y \rangle$ (dotted), which
obey exactly the relations $\langle v_x \rangle = (\sqrt{3}/2) \langle v \rangle$ and $\langle v_y \rangle = (1/2)\langle v \rangle$. Notice that the velocities converge exactly to $v_F/2$,
a value that can also be obtained analytically by making $k^0_yd
\rightarrow 0$ in Eq. (\ref{eqCompleteness}).

Henceforth, we will consider initial wave vectors $\vec{q}_0$
around the Dirac points 2 and 5 of Fig. \ref{fig:Honeycomb}(b),
namely
\begin{equation}\label{DiracPointsKKl}
K = \left(0,\frac{4\pi}{3\sqrt{3}a_0}\right) \quad and \quad K' =
\left(0,-\frac{4\pi}{3\sqrt{3}a_0}\right),
\end{equation}
respectively. This choice is very convenient, since the rotation
angles for these points are $\phi = \pi/2$ and $3\pi/2$,
respectively, so that the pseudo-spinor $[1,1]^T$ points to the
$y$ (-$y$) direction in the former (latter)
case. Hence, with this pseudo-spinor, wave packets in $K$ ($K'$)
will propagate with positive (negative) velocity in the vertical
zig-zag direction.

\subsection{External magnetic fields and strain}

\begin{figure}[!t]
\centerline{\includegraphics[width= \linewidth]{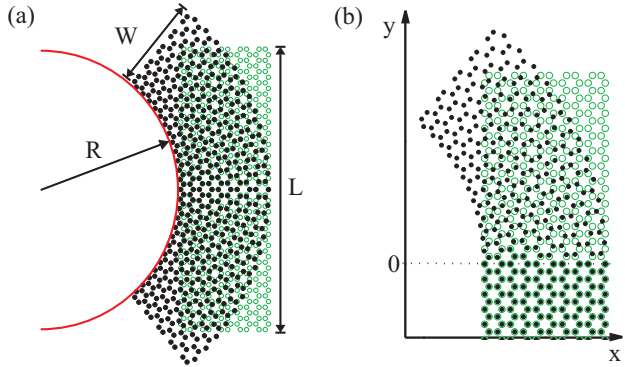}}
\caption{(Color online) (a) Sketch of the strained graphene sheet:
we consider a rectangular sample of width $W$ and height $L$, bent
into an arc of circle with inner radius $R$. The unstrained
graphene sheet is shown as open circles, for comparison. (b)
Strain-induced magnetic field barrier step, obtained by bending
the graphene lattice only in the $y \ge 0$ region. The number of
atoms was reduced in both figures, in comparison to the lattices
studied in this work, in order to improve the
visualization.}\label{fig:SketchARC}
\end{figure}

Recently,\cite{Guinea2} it was shown theoretically that bending a
graphene sheet into an arc of a circle produces a strong and
almost uniform pseudo-magnetic field profile. Figure
\ref{fig:SketchARC}(a) illustrates such a strained system, where
the rectangular graphene sample of width $W$ and height $L$ is
bent into an arc of a circle with inner radius $R$. As the
(pseudo) magnetic field points in the same direction (opposite
directions) at each $K$ and $K'$ points, \cite{Vozmediano} the
combination of both external and strain-induced magnetic field
effects provides a valley-dependent magnetic field. If one applies
the appropriate external magnetic field for some configuration of
the strained graphene, one can obtain an almost perfect
suppression of the effective magnetic field at one of the Dirac
cones, while the effective field in the other cone is enhanced.
This leads to a complicated system to be studied within the Dirac
approximation, since one has two completely different systems for
the $K$ and $K'$ valleys. Namely, Landau levels would be present
only around one of the cones (though one cannot expect a perfect
Landau level spectrum, since the strain-induced magnetic field is
not perfectly uniform in space), whereas in the other cone, the
usual continuum spectrum would be observed. This motivated us to
analyze the trajectories of a wave packet in such a system within
the TB model, where we do not need to include the pseudo-magnetic
fields artificially in the Dirac cones, since they appear
naturally when we consider the effect of the strain-induced
changes of the inter-site distances on the hopping energies, as
explained in the previous section.

In this subsection, we investigate the dynamics of a wave packet
with width $d = 200$ \AA\, and initial wave vector $k_x^0 = 0$ and
$k_y^0 = 0.02$ \AA\,$^{-1}$ around the Dirac points $K$ and $K'$
of Eq. (\ref{DiracPointsKKl}) in the presence of external and
strain-induced magnetic field barrier steps.  As in the $K'$
valley the pseudo-spinor $[1,1]^T$ is polarized in the negative
$y$-direction of the graphene lattice, we choose $[1,-1]^T$ for
this case, so that a wave packet in this valley will also
propagate in the positive $y$-direction. In order to obtain a
pseudo-magnetic field barrier step, we consider that the graphene
layer is strained only in the $y \ge 0$ region, as sketched in
Fig. \ref{fig:SketchARC}(b). We also consider an external magnetic
field $\vec{B} = B\Theta (y)\hat{z}$, where $\Theta(y)$ is the
Heaviside step function, which leads to a magnetic barrier step
for $y \ge 0$, described by the vector potential $\vec{A} =
(-By\Theta(y),0,0)$. In order to avoid effects due to
zitterbewegung in the (pseudo) magnetic field region, the wave
packet starts at the position $x_0 = 0$, $y_0 = -420$ \AA\,, so
that it can travel for some time in the magnetic field-free region
$y < 0$ until its velocity converge to a time independent value.
\begin{figure}[!h]
\centerline{\includegraphics[width=\linewidth]{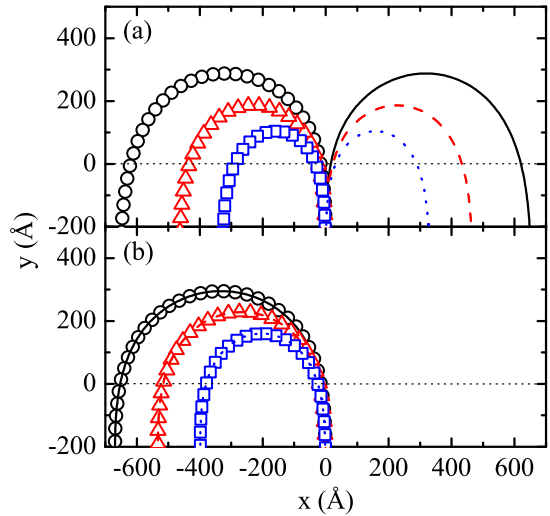}}
\caption{(Color online) Trajectories of the wave packet in the
$x-y$ plane, obtained by the TB method for such a system, for
initial momentum $k_y^0 = 0.02$ \AA\,$^{-1}$ around $K$ (symbols)
and $K'$ (curves) points, for (a) non-strained graphene with
magnetic barrier height $B = 5$ T (solid, circles), 7 T (dashed,
triangles) and 10 T (dotted, squares), and for (b) a graphene
sheet bent into an arc of circle with radius $R = 1$ $\mu$m
(solid, circles), 0.8 $\mu$m (dashed, triangles) and 0.6 $\mu$m
(dotted, squares), considering $B = 0$ T. In (b), symbols and
curves coincide for each value of $R$.}\label{fig:filterscheme}
\end{figure}

The influence of the external and strain-induced magnetic field
barriers on the trajectories of the wave packet are analyzed
separately in Fig. \ref{fig:filterscheme}, which shows the
trajectory of the centroid of the wave packets in $K$ (symbols)
and $K'$ (curves) points, calculated as $\langle r \rangle =
(\langle x \rangle, \langle y \rangle)$, (a) in a non-strained
graphene sheet with magnetic field barriers $B = 5$ T (solid,
circles), 7 T (dashed, triangles) and 10 T (dotted, squares) and
(b) in a strained graphene sheet with radius $R = 1$ $\mu$m
(solid, circles), 0.8 $\mu$m (dashed, triangles) and 0.6 $\mu$m
(dotted, squares). All the trajectories form semi-circles in the
$y \ge 0$ region, which is due to the Lorentz force produced by
the (pseudo) magnetic field. As the external magnetic field
(radius of the strained region) increases (decreases), the radii
of these semi-circular trajectories are reduced, since a higher
(pseudo) magnetic field produces a stronger modulus of the Lorentz
force. Notice that the radii of trajectories in the external and
pseudo-magnetic fields cases are comparable, which means that for
radii $R = 1$ $\mu$m - 0.6 $\mu$m of the strained graphene, the
generated pseudo-magnetic field is also within $\approx$ 5 T and
10 T. Indeed, the strain induced pseudo-magnetic field
distribution for the bend graphene ribbon is given by
\cite{Guinea2}
\begin{eqnarray}
B_S(x,y) = - 4c \frac{\beta\Phi_0}{aL}
\arcsin\left(\frac{L}{2R}\right)\cos\left[\frac{2x}{L}
\arcsin\left(\frac{L}{2R}\right)\right] \nonumber \\
\times\left[1 - \frac{R + y}{L}
\arcsin\left(\frac{L}{2R}\right)\right],\hspace{0.5 cm}
\label{strainguinea}
\end{eqnarray}
where $\beta \approx 2$ and $c$ is a dimensionless constant which
depends on the details of the atomic displacements. \cite{Guinea}
Considering $L/R \rightarrow 0$ in Eq. (\ref{strainguinea}) the
pseudo-magnetic field can be approximated as $B_S \approx
-c\beta\Phi_0\big/aR = \omega/R$. Using the value $\omega \approx
4.5 \times 10^4$ T\AA\, estimated numerically in Ref.
\onlinecite{Low}, one obtains pseudo-magnetic fields within
$B_S\approx 4.5$ T - 7.5 T for $R = 1$ $\mu$m - 0.6 $\mu$m, which
are of the same order of magnitude as the external magnetic fields
that we considered. For the external magnetic field barrier, the
trajectories of wave packets in $K$ and $K'$ points form circles
in opposite directions, as shown in Fig.
\ref{fig:filterscheme}(a), which is reasonable, since these
packets have opposite momentum, which causes a sign change in the
Lorentz force. Conversely, considering the strain-induced magnetic
barrier illustrated in Fig. \ref{fig:SketchARC}(b), the
trajectories of wave packets in $K$ and $K'$ curve in the same
direction, since, although their momenta have opposite signs, the
pseudo-magnetic fields also point in opposite directions at each
Dirac cone $K$ and $K'$.

\subsection{Strain induced valley filter}

\begin{figure}[!t]
\centerline{\includegraphics[width=\linewidth]{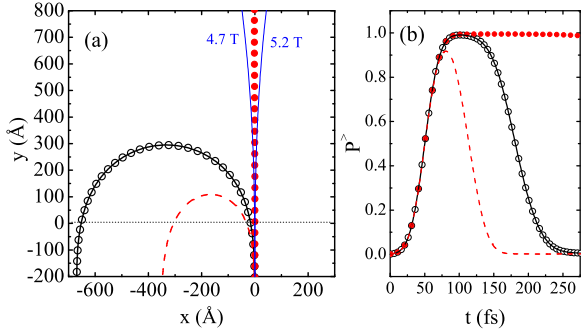}}
\caption{(Color online)(a) Trajectories on the $x-y$ plane for
wave packets with initial momentum $k_y^0 = 0.02$ \AA\,$^{-1}$
around $K$ (symbols) and $K'$ (curves) points, considering a
graphene sheet bent into an arc of circle with radius $R = 1$
$\mu$m and an external magnetic field $B = 0$ T (open, solid) and
4.9 T (full, dashed). The thin solid curves show the results for
two other magnetic field intensities for the $K$ packet. (b)
Probability of finding the particle in $y \ge 0$ as a function of
time, for wave packets with the same configurations as in
(a).}\label{fig:filterResult}
\end{figure}

Let us consider the strained sample in Fig. \ref{fig:SketchARC}(b)
with $R = 1$ $\mu$m. By comparing the radius of the semi-circular
trajectory of the wave packet in such a system with those obtained
for different intensities of the external magnetic field barrier,
one obtains the strain-induced magnetic field for this value of
$R$ as $\approx 4.9$ T. Figure \ref{fig:filterResult}(a) shows the
trajectories in the $x-y$ plane of the centroid of the wave
packets in a system where we combine a $R = 1$ $\mu$m strain for
$y \ge 0$ with an external magnetic field barrier $B = 0$ T
(solid, open) and 4.9 T (dashed, full), for wave packets in the
$K$ (symbols) and $K'$ (curves) Dirac points. In the absence of
the external magnetic field, both the $K$ and $K'$ packets exhibit
the same semi-circular trajectory, as discussed earlier. However,
when we combine the effect of the strain-induced and external
magnetic field barriers, the wave packet in $K'$ undergoes a
stronger Lorentz force and is readily reflected, whereas the one
in the $K$ point performs a practically straight trajectory, as if
this packet is not influenced by any Lorentz force. This is a
consequence of the fact that combining the effects of a
pseudo-magnetic field produced by a $R = 1\mu$m strain and a $B =
4.9$ T external magnetic field produces a stronger magnetic field
in the $K'$ point, while in the $K$ point these fields
equilibrate, producing a practically magnetic field-free region
for particles in this cone. In this situation, the system works as
a valley filter, where only wave packets in the $K$ Dirac cone are
allowed to pass through the strained region, whereas the wave
packets in $K'$ are reflected. The results for the wave packet in
$K$ for two other values of the external magnetic field are shown
as thin solid lines, showing that within a range of $\Delta B =
\pm 0.2$ T around $B = 4.9$ T, which is a reasonable range for
magnetic field intensities in experiments, only a weak Lorentz
force is observed and the valley filter works fine.

The results of Fig. \ref{fig:filterResult} are obtained for both
external and pseudo-magnetic field barriers starting at the same
position $y = 0$. It is straightforward to verify that if there is
a mismatch between the starting points of the strained and
external field regions, some deviations will occur in the
trajectories of the wave packets but, provided the length of the
mismatch is much smaller than the magnetic length, the filtering
effect is still stable. As an example, a 30~\AA\, mismatch between
the external and pseudo-magnetic field barriers in the system
analyzed in Fig. \ref{fig:filterResult} would produce a $\approx$
5$^\circ$ deviation in the otherwise vertical trajectory of the
wave packet in $K$, whereas the wave packet in $K'$ is still
readily reflected by the combination of magnetic fields in the
filter region.

The probability $P^>$ of finding the particle in the strained $y \ge 0$
region, calculated as
\begin{equation}
P^> (t)= \sum_{n^>}\sum_{m}|\Psi_{n,m}^t|^2,
\end{equation}
where $n^>$ represents the lines of atomic sites with $y \ge 0$,
is shown as a function of time in Fig. \ref{fig:filterResult}(b).
In the $B = 0$ T case, both wave packets in $K$ (open circles) and
$K'$ (solid) stay in the strained $y \ge 0$ region until $t
\approx 300$ fs, when they turn back into the $y < 0$ region,
reflected by the Lorentz force induced by the strain. However, for
$B = 4.9$ T, $P^>$ already approaches zero at $t \approx 175$ fs
for the packet in $K'$ (dashed), whereas for $K$ (full circles),
it remains close to 1 even for large $t$.
\begin{figure}[!b]
\centerline{\includegraphics[width= 0.4\textwidth]{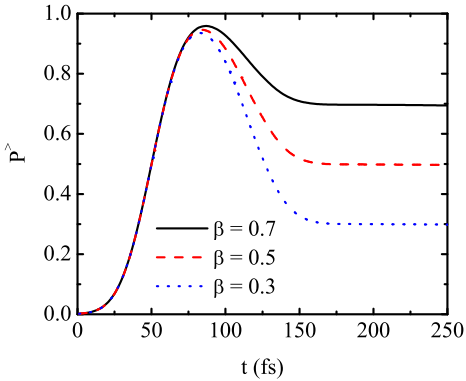}}
\caption{(Color online) Probability of finding the electron in the
filter region $y \ge 0$, for an initial wave packet given by a
combination of Gaussian distributions around both $K$ and $K'$
Dirac points, for three different values of the $K$-component of
the wave packet $\beta$.}\label{fig:filterEfficiency}
\end{figure}

The efficiency of the proposed valley filter is confirmed by Fig.
\ref{fig:filterEfficiency}, where we present $P^>$ as a function
of time for initial wave packets given by a combination of
Gaussians around the $K$ and $K'$ points in Eq.
(\ref{DiracPointsKKl}):
\begin{equation}
\Psi = \sqrt{\alpha} \Psi_{K'} + \sqrt{\beta} \Psi_{K},
\end{equation}
where $\Psi_{K(K')}$ is the Gaussian wave packet in Eq.
(\ref{initial}) with momentum $\vec{q}$ around the $K (K')$ Dirac
point. The results are presented for three different values of
$\beta$, where one can easily see that the probability of finding
the packet in the strained region exhibits a peak at $t \approx
80$ fs but, as the $K'$ part of the packet is reflected by the
magnetic barrier, this probability decays, reaching exactly $P^> =
\beta$ for large $t$. Such a system proves to be a perfect valley
filter, which is able to reflect all the components of the
incoming packet that are in the $K'$ point and transmit a wave
packet that is composed only of particles with momentum in the
vicinity of $K$.

We point out that when a wave packet reaches the edges of a
graphene nanoribbon, it can be scattered to a different Dirac
cone. \cite{Chen} Consequently, the efficiency of the valley
filter could be compromised if one considers a thin nanoribbon, so
that the filtered wave packet could still reach its boundaries and
scatter back to the other valley. In order to avoid such an
effect, we have considered a wide nanoribbon, so that for the time
intervals we study in this work, boundary effects are not
significant.

\subsection{External and pseudo magnetic field effects on the zitterbewegung}

In a previous paper, Rusin and Zawadzki \cite{Rusin1} used the
Dirac Hamiltonian for graphene to show that the zitterbewegung,
which is transient for $B$ = 0, as discussed earlier, is permanent
for $B \ne 0$. Furthermore, the authors showed that for a Gaussian
wave packet, the time evolution of the average value of the
current is different in $x$ and $y$ directions, which they explain
as due to the fact that the Dirac Hamiltonian is not symmetric in
the momenta $\hat{p}_x$ and $\hat{p}_y$. Although the same authors
say in their subsequent paper \cite{Rusin2} that this effect is
unphysical, because it violates the rotational symmetry of the
$x-y$ graphene plane, we believe this result is still physical:
one must remember that the Dirac model of graphene comes from the
tight-binding approach for this system, which describes a
honeycomb lattice of atoms that is not symmetric in the $x-y$
plane by definition, exhibiting C$_{6v}$ symmetry, as mentioned in
previous section. For each $K$ and $K'$ point, the coordinates
$\textsf{x}$ and $\textsf{y}$ in the Dirac Hamiltonian represent
different directions in the real honeycomb lattice of graphene,
where for an infinite sample the $\textsf{x}$ ($\textsf{y}$)
coordinate in the Dirac equation is related to one of the zigzag
(armchair) directions of the real sample. In this subsection, we
use our TB model of graphene to extend the previous study of Rusin
and Zawadzki \cite{Rusin1} to different situations.

We now study the dynamics of a wave packet with width $d = 200$
\AA\,, pseudo-spinor $c_1 = 1$ and $c_2 = 1$ and initial wave
vector $k_x^0 = k_y^0 = 0$, i.e. exactly at one of the Dirac
points $K$ and $K'$ in Eq. (\ref{DiracPointsKKl}), in the presence
of an uniform applied external magnetic field $\vec{B} = B
\hat{z}$, instead of the magnetic field barrier step considered in
the previous subsection. We also consider a pseudo-magnetic field
obtained by bending the whole rectangular graphene sample into an
arc of a circle with radius $R$, as illustrated in Fig.
\ref{fig:SketchARC}(a). The radius of the circle is considered as
$R = 1$ $\mu$m, which produces a $\approx 4.9$ T pseudo-magnetic
field, as demonstrated in the previous subsection. Accordingly, we
consider the external uniform magnetic field as $B = 4.9$ T.

A few experimental techniques have been suggested in the
literature for the observation of zitterbewegung.
\cite{Schliemann, Gerritsma} An interesting one \cite{Rusin2} is
based on the fact that the wave packet $\Psi(\vec{r},t)$ exhibits
an electric dipole moment $\vec{D}(t) = \langle
\Psi(\vec{r},t)|\vec{r}|\Psi(\vec{r},t)\rangle$ and, consequently,
the zitterbewegung yields an oscillation of this dipole moment,
which is a source of electromagnetic radiation, described
classically \cite{Jackson} by the equation
\begin{equation}
\vec{\varepsilon}(t) =
\frac{d^2\vec{D}(t)}{dt^2}\frac{\sin{\Phi}}{4\pi \epsilon_0s},
\end{equation}
where $s$ and $\Phi$ are the distance and angle of observation,
respectively, and $\epsilon_0$ is the vacuum permittivity.
\begin{figure}[!t]
\centerline{\includegraphics[width= 0.4\textwidth]{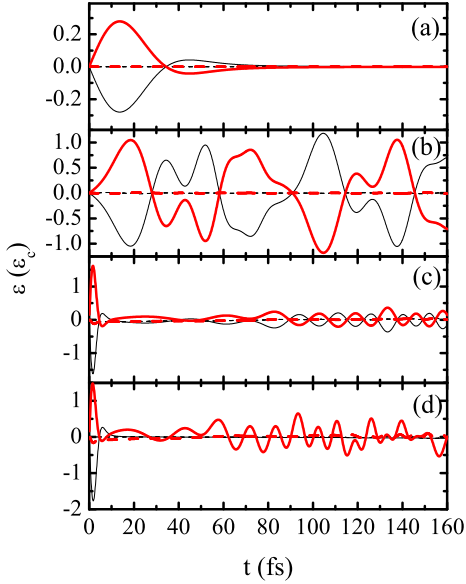}}
\caption{(Color online) Electromagnetic dipole radiation as a
function of time for wave packets with initial pseudo-spinor
$[1,1]^T$ at $K$ (thin curves) and $K'$ (thick curves),
considering a graphene sheet (a) in the absence of strain and
magnetic fields, (b) under an uniformly applied magnetic field $B
= 4.9$ T, (c) bent into an arc of circle with radius $R = 1 \mu$m
(see Fig. \ref{fig:SketchARC}) and (d) with both the uniform
magnetic field $B = 4.9$ T and the $R = 1 \mu$m bending. Solid
(dashed) curves are the results for the $\varepsilon_y$
($\varepsilon_x$) component.}\label{fig:Efield}
\end{figure}

Figure \ref{fig:Efield} shows the effects of external and
strain-induced magnetic fields on the electric field radiation
produced by the zitterbewegung, written in units of $\varepsilon_c
= \sin{\Phi}\big/4\pi \epsilon_0 s$. Only weak oscillations are
expected in the armchair ($x$) direction, since the pseudo-spin
$[1,1]^T$ points in the $ \textsf{x}$-direction in the Dirac model
which, as mentioned earlier, is related to the zigzag ($y$)
direction of the honeycomb lattice. Indeed, in Fig.
\ref{fig:Efield}(a-d), the $x$-component of the electric field
(dashed) is always close to zero. In Fig. \ref{fig:Efield}(a), we
present the results in the absence of strain and magnetic fields,
for comparison. In this case, the oscillations on the electric
field are suppressed for larger time, which is due to the
transient character of the zitterbewegung. Besides, the results
for $\varepsilon_y$ (solid) in the $K$ (thin curves) and $K'$
(thick curves) points have opposite signs, since for these points
the axis of the Dirac cones are rotated by angles which differ by
$\pi/2$ difference. In Fig. \ref{fig:Efield}(b), the uniformly
applied magnetic field $B = 4.9$ T in an unstrained sample leads
to persistent oscillations in $\varepsilon_y$, which is related to
the discrete Landau level spectrum created by this field. Each
Landau level that is populated by the Gaussian distribution
contributes with a different frequency. \cite{Rusin1} Figure
\ref{fig:Efield}(c) shows that such a persistent behavior is also
obtained by bending the graphene sheet into an arc of circle with
radius $R = 1\mu$m, which produces a pseudo-magnetic field of the
same order of magnitude. Notice that the amplitude of oscillations
in this case is four times smaller than those found in Fig.
\ref{fig:Efield}(b) for the unstrained sample under an external
magnetic field. In fact, these two cases are not expected to
produce the same zitterbewegung, because, although both samples
exhibit approximately the same magnitude of magnetic field, the
strained sample has not only the pseudo-magnetic field, but also a
different distribution of atomic sites. Thus, in the strained
case, there is an additional change in the direction of the
pseudo-spin polarization as the wave packet drifts, due to the
lattice distortion itself. As we have demonstrated in Sec.~III~A,
the zitterbewegung strongly depends on the pseudo-spin
polarization and hence, the different interplay between the
strained atomic sites and the initial pseudo-spin polarization
produces a different zitterbewegung for the strained case, as
compared to the one of the unstrained sample under an external
field.

In Fig. \ref{fig:Efield}(d), we combine the effects of the $R =
1$~$\mu$m strain and $B = 4.9$ T external field to produce a
system where the magnetic field is practically zero in the $K$
point, but is non-zero in the $K'$ point, so that only the packet
in the $K'$ point exhibits persistent oscillations. For the $K$
point, the external field compensates only the effect of the
pseudo-magnetic field, namely, the persistent zitterbewegung, but
it does not remove the effect of the lattice distortion. As a
result, comparing the results for $K$ (thin curves) in Figs.
\ref{fig:Efield}(a) and (d), one observes that the oscillations
are transient in both cases, since there is no effective magnetic
field, but they still exhibit a different oscillation profile at
small $t$, due to the lattice distortion in the latter case, which
is absent in the former.

\section{Conclusion}

We presented a study of the dynamics of Gaussian wave packets in
graphene under external and strain-induced magnetic fields, where
the latter is obtained by bending the graphene sheet into an arc
of a circle. The dependence of the zitterbewegung on the initial
pseudo-spin of the wave packet is investigated, and the results
obtained by means of the tight-binding model and the Dirac
equation are compared. We demonstrate that the combination of the
pseudo-magnetic field, induced by bending the graphene sheet,
along with an external magnetic field with appropriate strength
can be used as an efficient valley filter. An incoming wave packet
composed of momenta around the $K$ and $K'$ Dirac points is
scattered such that all its components in one of the Dirac cones
undergoes a strong Lorentz force and are readily reflected, while
the components in the other cone are allowed to pass through the
device with only small distortions in their trajectory, due to the
very weak residual Lorentz force.

Our results also show that in the absence of external or
strain-induced magnetic fields, the zitterbewegung is a transient
effect, whereas in the presence of any of these fields, the
oscillations persist in time. In a strained sample under an
external magnetic field with the appropriate strength, the
effective magnetic field in one of the Dirac cones is enhanced,
whereas in the other cone it is practically cancelled. In this
situation, a permanent zitterbewegung is observed only for wave
packets in one of the Dirac cones. The wave packet oscillations
produce electric field radiation, which can be detected
experimentally.

Finally, we believe the present work contributes to a better
understanding of the relation between the results obtained from TB
and Dirac approaches for graphene and those to be observed in
future experiments on strained graphene-based structures.

\acknowledgements This work was financially supported by CNPq,
under contract NanoBioEstruturas 555183/2005-0,
PRONEX/CNPq/FUNCAP, CAPES, the Bilateral programme between
Flanders and Brazil, the Belgian Science Policy (IAP) and the
Flemish Science Foundation (FWO-Vl).

\end{document}